\documentclass[conference,a4]{IEEEtran}
\IEEEoverridecommandlockouts

\IEEEoverridecommandlockouts
\usepackage{cite}
\usepackage{amsmath,amssymb,amsfonts}
\usepackage{algorithmic}
\usepackage{graphicx}
\usepackage{textcomp}
\usepackage{xcolor}
\usepackage[font={footnotesize}]{caption}
\usepackage{subcaption}
\usepackage{cite}
\usepackage{balance}
\usepackage{amsmath,amssymb,amsfonts}
\usepackage{algorithmic}
\usepackage{graphicx}
\usepackage{textcomp}
\usepackage{xcolor}
\usepackage{tikz}
\usepackage{pgfplots} 
\usepackage{tikzscale}
\usepackage{pgfplots}
\pgfplotsset{compat=newest}
\pgfplotsset{plot coordinates/math parser=false}
\newlength\figureheight
\newlength\figurewidth
\def\BibTeX{{\rm B\kern-.05em{\sc i\kern-.025em b}\kern-.08em
    T\kern-.1667em\lower.7ex\hbox{E}\kern-.125emX}}
\begin{document}

\title{Bi-Static Sensing for Near-Field RIS Localization \\
\author{Reza Ghazalian\IEEEauthorrefmark{1}, Kamran Keykhosravi\IEEEauthorrefmark{2}, Hui Chen\IEEEauthorrefmark{2}, Henk Wymeersch\IEEEauthorrefmark{2}, Riku J{\"a}ntti\IEEEauthorrefmark{1}\\
\IEEEauthorrefmark{1}Aalto University, Finland, 
\IEEEauthorrefmark{2}Chalmers University of Technology, Sweden \\
E-mail: \{reza.ghazalian, riku.jantti\}@aalto.fi, \{kamrank, hui.chen, henkw\}@chalmers.se
}
}


\maketitle

\begin{abstract}
We address the localization of a  reconfigurable intelligent surface (RIS) for a single-input single-output multi-carrier system using bi-static sensing between a fixed transmitter and a fixed receiver. Due to the deployment of RISs with a large dimension,
near-field (NF) scenarios are likely to occur, especially for indoor applications, and are the focus of this work. 
We first derive the Cram\'er-Rao bounds (CRBs) on the estimation error of the RIS position and orientation and the time of arrival (TOA) for the path transmitter-RIS-receiver. We propose a multi-stage  low-complexity estimator for RIS localization purposes. In this proposed estimator, we first perform a  line search to estimate the TOA. Then, we use the far-field approximation of the NF signal model to implicitly estimate the angle of arrival and the angle of departure at the RIS center. 
Finally, the RIS position and orientation estimate are refined via a quasi-Newton method. Simulation results reveal that the proposed estimator can attain the CRBs. We also investigate the effects of several influential factors on the accuracy of the proposed estimator like the RIS size, transmitted power, system bandwidth, and RIS position and orientation.

\end{abstract}


\section{Introduction}
Reconfigurable intelligent surfaces (RISs) consist of controllable elements that can alter the characteristics of an electromagnetic (EM) wave, such as phase, amplitude, frequency, or polarization \cite{wymeersch2020radio,basar2019wireless,yuan2021frequency}. By properly configuring the element coefficients, RISs provide controllable communication channels to support communication and localization services for wireless devices \cite{strinati2021reconfigurable}. With the properties of being energy-efficient, lightweight, easy to deploy, and compatible with the existing wireless infrastructures, RISs are expected to be the vital enabler for the sixth generation (6G) communication systems \cite{basar2020reconfigurable,chen2021tutorial}.


 RIS-aided communication and localization have been intensely studied in recent years, see, e.g., \cite{xing2021location,keykhosravi2021siso,abu2021near,huang2022near, he2020adaptive}. In the RIS-assisted communications context, for instance, a location-aware beamforming design in the far-field (FF) region was proposed in  \cite{xing2021location}, where authors  established the geometrical channel model based on the user's 3D location. In the localization context, the 3D position of a user with the aid of RIS has been estimated in \cite{keykhosravi2021siso}, where the authors have considered the FF model. For 3D user localization in the near-field (NF) region of an RIS acting as a lens, a low-complexity estimator has been proposed in \cite{abu2021near}. For 2D RIS-aided user positioning, the RIS-assisted received signal strength (RSS)-based localization algorithms have been proposed in \cite{huang2022near}. In \cite{he2020adaptive}, the linear RIS phase profile has been designed for both communications and positioning systems for the FF region in a 2D scenario. 
According to the recent works in RIS-aided localization systems, it can be realized that the location and orientation of the RIS need to be known. That is because RIS is one of the reference points in these systems \cite{wymeersch2020radio}. In the context of RIS-aided communication  systems, knowing RIS location can facilitate the location-based RIS phase profile optimization \cite{hu2020location}. However, the \emph{location and the orientation of the RIS were assumed to be known} in all previous studies, which is non-trivial to obtain in practice.  The problem of RIS localization (i.e., determining the position of an RIS) can be seen as a bi-static sensing problem, and was previously tackled in \cite{keykhosravi2021semi} under FF conditions. However, the RIS orientation could not be determined, given the time of arrival (TOA) measurements. When the RIS is large, devices in most RIS-aided systems will be in  the NF, which makes the RIS localization problem more challenging than in \cite{keykhosravi2021semi}.

In this paper, we aim at estimating the location and orientation of an RIS in a NF scenario with a single-antenna receiver (RX), where a multi-carrier (MC) single-antenna transmitter (TX) sends signals to the RX via the linear RIS. 
Our main contributions are as follows: (i) we formulate the RIS localization problem in the NF region for a single-input and single-output (SISO)-MC system with a blocked line of sight (LOS) path, (ii) we derive a Fisher information matrix (FIM) for RIS location and orientation estimate, (iii) we propose a multi-stage low-complexity RIS localization algorithm and evaluate it by comparing with the theoretical error bound; Moreover, we use FF approximation of the NF observed signals to earn information about the angle-of-arrival (AOA) and the angle-of-departure (AOD) at the RIS.  

\begin{figure}[t]
\centering
\includegraphics[width=0.75\columnwidth]{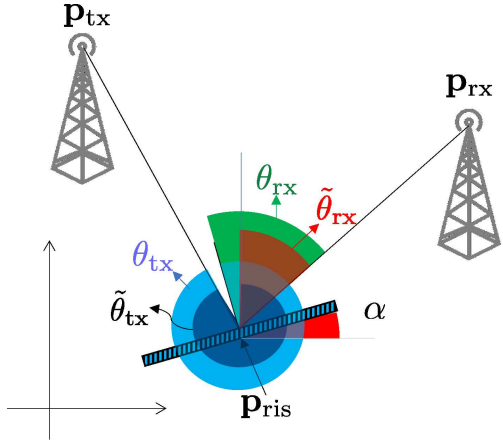}
\caption{System setup, comprising single-antenna TX and RX, as well as an RIS with unknown location and orientation. $\tilde{\theta}_\text{tx}$ and  $\tilde{\theta}_\text{rx}$ represent the AOA and AOD at the RIS center when $\alpha=0$, respectively.} 
\label{System Setup}
\end{figure}
\subsubsection*{Notation}
Vectors and matrices are indicated by lowercase and uppercase bold letters, respectively. The element in the $i$th row and
$j$th column of matrix $\mathbf{A}$ is specified by $\left[\mathbf{A}\right]_{i,j}$. The sub-index
$i:j$ determines all the elements between $i$ and $j$. The complex conjugate, Hermitian, and transpose
operators are represented by $\left( .\right)^*$, $\left( .\right)^{\mathsf{H}}$, and $\left( .\right)^\top$, respectively. $\Vert.\Vert$ calculates the norm of vectors or Frobenius norm of matrices. By $\odot$, we indicate element-wise product.  $\jmath=\sqrt{-1}$ and $\mathbf{1}_K$ is a column vector comprising all ones with length $K$. The function $\text{atan2}(y,x)$ is the four-quadrant inverse tangent function.
\section{System Model}
In this section, we describe the considered wireless system for
2D\footnote{Our study is limited to 2D, given the inherent difficulty of RIS localization. The 3D problem is left for future work.} RIS localization and elaborate the signal model.
\subsection{System Setup}
We consider a wireless system consisting of a single-antenna TX, a single-antenna RX, and an $M$-elements linear RIS in a two-dimensional Cartesian coordinate system (global reference) as shown in Fig.~\ref{System Setup}. The RX and TX are located at NF region\footnote{NF region, which is usually defined as the range smaller than the Fraunhofer distance $d_F = 2D^2/\lambda$ \cite{sherman1962properties}. Here, $D$ and $\lambda$ are the dimension of the RIS and the wavelength at the carrier frequency, respectively.} of the RIS with the LOS path blocked.\footnote{This is not a limiting assumption, but only invoked for notational convenience. When the LOS path is present, it can be separated from the RIS path as in \cite{keykhosravi2021semi}. When the TX and RX are synchronized, the LOS path does not convey any information.} We also assume that TX and RX are  synchronized (i.e., known clock bias) with known position ($\mathbf{p}_{\text{tx}}$ and $\mathbf{p}_{\text{rx}}$) (e.g., RX  as calibration agent). 

With a rotation angle $\alpha$, the global position of $m$th RIS element can be expressed as a function of the RIS center $\mathbf{p}_{\text{ris}}$: 
\begin{equation}\label{eq:p-global}
    \mathbf{p}_{m} = \mathbf{p}_{\text{ris}} + \mathbf{R}_{\alpha} \left(m-\frac{M-1}{2} \right)\Delta,
\end{equation}
where $\Delta$ is the inter-element space (e.g., ${\lambda}/{2}$) and  $\mathbf{R}_\alpha$ is the 2D rotation matrix as
\begin{equation}
\mathbf{R}_\alpha = \begin{bmatrix}
 \cos{(\,\alpha)\,} & -\sin{(\,\alpha})\,\\
  \sin{(\,\alpha)\,} & \cos{(\,\alpha})\, \\
\end{bmatrix}.
\end{equation}
The aim of this work is to estimate the orientation ($\alpha$) and the center position of the RIS ($\mathbf{p}_{\text{ris}}$).




\subsection{Signal Model}
We consider a MC scenario, where TX transmits $T$ orthogonal frequency division multiplexing (OFDM) symbols over time with $N_c$ sub-carries for RIS localization purposes. Without loss of generality, we assume that all the transmitted symbols are equal to one. It is also assumed that RIS changes the phase of the incident wave through a random profile over time (denoted by $t$). The RIS phase profile at time $t$ is shown by the vector $\boldsymbol{\gamma}(t) \in \mathbb{C}^{M\times1}$, where $|[\boldsymbol{\gamma}(t)]_m|= 1$ for all $m$.  Based on these assumptions, we can write the received signal over time and sub-carrier frequencies as the following $N_c\times T$ matrix 
\begin{align}\label{eqn:observation}
     \mathbf{Y}=
\rho e^{\jmath \phi} \sqrt{P_t} \mathbf{d}\left( \tau\right) \mathbf{b}^\top \boldsymbol{\Gamma}+ \mathbf{W}, 
\end{align}
where $\rho e^{\jmath \phi}$ is an unknown complex gain, modeled with $\phi \sim \mathcal{U}[0,2\pi)$ and \cite{tang2020wireless}
\begin{align}
    \rho = \sqrt{\frac{G_{t} G_{r} A \lambda^2}{64\pi^3}}\frac{1}{ \|\mathbf{p}_{\text{tx}}-\mathbf{p}_{\text{ris}}\| \|\mathbf{p}_{\text{rx}}-\mathbf{p}_{\text{ris}}\|},
\end{align}
where $A=\frac{\Delta^2}{4}$ denotes the RIS element's surface area and $G_{t}$ and $G_{r}$ are the antenna gain at the TX and RX, respectively. {$P_t$ denotes the transmitted power.} The matrix $\mathbf{\Gamma}$ contains all of the RIS phase profiles over time, that is $ \mathbf{\Gamma}= \begin{bmatrix}
\boldsymbol{\gamma}(0), \dots ,\boldsymbol{\gamma} (T)
\end{bmatrix}$,  $\mathbf{W} \in \mathbb{C}^{N_c\times T}$ is the noise matrix containing zero-mean circularly-symmetric independent and identically distributed Gaussian elements with variance $\sigma^2$. We consider the effect of the delay on the signal with the  vector $\mathbf{d}(\tau)\in \mathbb{C}^{N_c\times1}$, which is defined as 
\begin{equation}
    \mathbf{d}(\tau) = [ \, 1, e^{-\jmath 2 \pi\tau \Delta f}, \dots ,e^{-\jmath 2 \pi\tau (N_c-1) \Delta f}  ] \,^\top,
\end{equation}
where $\Delta f$  is sub-carrier spacing and $\tau = \|\mathbf{p}_{\text{tx}}-\mathbf{p}_{\text{ris}}\|/c+ \|\mathbf{p}_{\text{rx}}-\mathbf{p}_{\text{ris}}\|/c$ for speed of light $c$. Finally, $\mathbf{b}=\mathbf{a}_t \odot \mathbf{a}_r$, where 
the vectors $\mathbf{a}_t$ and $\mathbf{a}_t$ are the NF array response vectors from the RIS to the TX and RX, respectively. The $m$th elements of these vectors are defined as  \cite{friedlander2019localization}
\begin{subequations} \label{eq:RISreponses}
\begin{equation}\label{eq:at}
 [\mathbf{a}_t]_m = e^{-\jmath \frac{2\pi}{\lambda} \left(R^{t}_{m}-R^{t}_o\right)} ,
\end{equation}
\begin{equation}\label{eq:ar}
 [\mathbf{a}_r]_m = e^{-\jmath \frac{2\pi}{\lambda} \left(R^{r}_{m}-R^{r}_o\right)}, \end{equation}
\end{subequations}
where 
$R^{t}_{m}=\Vert \mathbf{p}_m -  \mathbf{p}_{\text{tx}}\Vert$,  $R^{r}_{m}=\Vert \mathbf{p}_m -  \mathbf{p}_{\text{rx}}\Vert$, $R^{t}_o=\Vert \mathbf{p}_{\text{ris}} -  \mathbf{p}_{\text{tx}}\Vert$, $R^{r}_o=\Vert \mathbf{p}_{\text{ris}} -  \mathbf{p}_{\text{rx}}\Vert$. Here, we assume that the wavelength remains relatively constant over the transmission bandwidth (BW), so that beam squint effects can be ignored. 


\section{Fisher Information Analysis}

Introducing the noiseless part of observation given in \eqref{eqn:observation} as  $\boldsymbol{\mu} =  \rho e^{\jmath\phi}\sqrt{P_t}
    \mathbf{d}(\,\tau)\,\mathbf{b}^\top  \mathbf{\Gamma}$ , and
 the $6\times1$ vector of unknowns $\boldsymbol{\eta} = [\rho \quad \phi \quad \tau \quad \mathbf{p}_{\text{ris}}^\top \quad \alpha ]^\top$, and $\boldsymbol{\zeta} = \left [\rho \quad \phi \quad \mathbf{p}_{\text{ris}}^\top \quad \alpha \right]^\top$,  the FIM $\mathbf{J}(\boldsymbol{\eta}) \in  \mathbb{R}^{6\times 6}$ is defined as \cite[Sec. 3.9]{kay1993fundamentals}
\begin{align}\label{eqn:FIM WB}
     \mathbf{J}(\boldsymbol{\eta})= \frac{2}{\sigma^2} \sum_{t=1}^T\sum_{n_c=1}^{N_c} \Re \Bigg\{\frac{ \partial [\boldsymbol{\mu}]_{t,n_c}}{\partial\boldsymbol{\eta}} \left( \frac{ \partial [\boldsymbol{\mu}]_{t,n_c}}{\partial\boldsymbol{\eta}}\, \right)^{\mathsf{H}}\, \Bigg\}.
\end{align}
 Using \eqref{eqn:FIM WB}, we can calculate the Cram\'er-Rao bound (CRB) of the $\tau$, i.e., the time error bound (TEB) as
 \begin{align}\label{eqn:TEB}
     \sqrt{\mathbb{E}\left [\left( \tau -\hat{\tau} \right)^2  \right]}\geq \text{TEB}\triangleq  \sqrt{\left[ \mathbf{J}(\boldsymbol{\eta})^{-1}\right]_{3,3}} \quad,
\end{align}
 where $\hat{\tau}$ is the estimate of the true $\tau$. Since there is a relationship between $\tau$ and $\mathbf{p}_{\text{ris}}$, the FIM of position and orientation $\mathbf{J}\left( \boldsymbol{\zeta}\right)$ can be derived as  $\mathbf{J}\left( \boldsymbol{\zeta}\right) = \mathbf{T}^\top \mathbf{J}\left( \boldsymbol{\eta}\right) \mathbf{T}$, where $\mathbf{T}\in \mathbb{R}^{6\times5}$ is a Jacobian matrix \cite[Eq. (3.30)]{kay1993fundamentals}. Its element on the $\ell$th row and the $q$th column is defined
\begin{align}\label{eqn:Jacobian}
     \left[\mathbf{T}\right]_{\ell,q} = \frac{\partial \left[\boldsymbol{\eta}\right]_{\ell}}{\partial\left[\boldsymbol{\zeta}\right]_{q}} \quad.
\end{align}

Considering \eqref{eqn:FIM WB} and \eqref{eqn:Jacobian}, the position error bound (PEB) and orientation (EOB) can be written
as
\begin{subequations}
\begin{align}\label{eq:PEB}
 \sqrt{\mathbb{E}\left [ \Vert\mathbf{p}_{\text{ris}} -\hat{\mathbf{p}}_{\text{ris}}\Vert^2   \right]}\geq \text{PEB}\triangleq  \sqrt{\mathrm{tr}(\left[ \mathbf{J}(\boldsymbol{\zeta})^{-1}\right]_{3:4,3:4})} \quad,
\end{align}
\begin{align}\label{eq:OEB}
 \sqrt{\mathbb{E}\left [ \left(\alpha -\hat{\alpha}\right)^2   \right]}\geq \text{OEB}\triangleq  \sqrt{\left[ \mathbf{J}(\boldsymbol{\zeta})^{-1}\right]_{5,5}} \quad, \end{align}
\end{subequations}
where $\hat{\mathbf{p}}_{\text{ris}}$ and $\hat{\alpha}$ are the estimates of the true RIS position and orientation, respectively. The matrices $\mathbf{J}$ and $\mathbf{T}$ are calculated in the Appendix.


\section{RIS Location and Orientation Estimator}
\subsection{Maximum Likelihood Estimator}
Based on the observation given in (\ref{eqn:observation}), the Maximum Likelihood (ML) estimator is defined as 
\begin{align}\label{eq:MLE}
  \left[\hat{g}_r,\hat{\alpha},\hat{\mathbf{p}}_{\text{ris}} \right] &=  \arg\max_{\alpha, \mathbf{p}_{\text{ris}}} f\left( \mathbf{Y} |g_r , \alpha, \mathbf{p}_{\text{ris}}  \right)\\ \nonumber
  &= \arg\min_{g_r, \alpha, \mathbf{p}_{\text{ris}}} \Vert \mathbf{Y} - g_r \sqrt{P_t} \mathbf{d}\left(\tau\right) \mathbf{b}^\top \boldsymbol{\Gamma}\Vert^2, \end{align}
where $g_r = \rho e^{\jmath\phi}$. To solve \eqref{eq:MLE}, we can use any gradient descent method (e.g., Newton method). However, the function described in \eqref{eq:MLE} is non-convex and has many  local optima  around the global optimum. Thus finding a suitable initial estimate for gradient descent methods is challenging. To tackle this issue, we propose a low-complexity estimator, which can provide such an initial estimate. 
\subsection{Low Complexity Estimator}
 In this section, we develop a low-complexity estimator for RIS orientation and position estimation. To this end, first, we estimate $\tau$. Then we use the FF approximation of the NF to gain information about the AOA and the AOD at the RIS. Based on these estimations, we can find the RIS position and orientation through a line search. Finally, we adopt this estimate as an initial guess for the quasi-Newton algorithm to solve the problem \eqref{eq:MLE}. We indeed use the quasi-Newton algorithm to refine the initial estimate. 

\subsubsection{Estimation of $\tau$}
For $\tau$ estimation, we take the IFFT with length $N_{F}$ of the columns of the matrix $\mathbf{Y}$, yielding $\mathbf{Z}=\mathbf{F} \mathbf{Y}$. We then obtain a coarse estimate of the TOA as 
\begin{align}
    \tilde{k} = \arg \max_{k} \Vert \mathbf{g}^\top_k \mathbf{Z}\Vert,
\end{align}
where $\mathbf{g}_k$ is vector
comprising $N_F$ zeros, except with one in the $k$th entry. 
Accordingly, one can obtain a refined estimate $\tau$ as follows \cite{keykhosravi2021semi}:
\begin{equation}
\label{eq:tau}
  \hat{\tau} = \frac{\tilde{k}}{N_{F}\Delta f}-\tilde{\delta},
\end{equation} 
where
\begin{align}
    \tilde{\delta}= \arg \max_{\delta \in \left[0 , 1/\left(N_F \Delta f\right)\right]} \Vert \mathbf{g}^\top \mathbf{Z} \ \odot \left(\mathbf{d}\left(\delta \right)\mathbf{1}_T^\top\right)\Vert.
\end{align}
\subsubsection{Estimating AOA and AOD at the center of RIS}
We first remove the effect of $\tau$ from $\mathbf{Y}$, i.e.,
\begin{align}\label{eq:remove_d}
  \mathbf{Y}_r &= \mathbf{Y} \odot \left(\mathbf{d}\left( -\tau\right) \mathbf{1}_T^\top\right)\\
  \nonumber
  & = g_r \sqrt{P_t} \mathbf{1}_{N_c} \mathbf{b}^\top \boldsymbol{\Gamma} + \mathbf{W}_d ,
\end{align}
where $\mathbf{W}_d\triangleq \mathbf{W} \odot \left(\mathbf{d}\left( -\tau\right) \mathbf{1}_T^\top\right) $. 
Next, we sum the signal $\mathbf{Y}_r$ across the sub-carriers to obtain 
\begin{align}\label{eq:sum-subcar}
\mathbf{y}_r & = \mathbf{Y}_r^\top \mathbf{1}_{N_c} \Rightarrow
\mathbf{y}_r = N_c g_r \sqrt{P_t} \boldsymbol{\Gamma}^\top \mathbf{b} + \mathbf{w}_t ,
\end{align}
where $\mathbf{w}_t \triangleq \mathbf{W}^\top \mathbf{1}_{N_c}$. 
In the NF regime, AOA and AOD can be defined for each array element. However, estimating all these angles for all elements is not possible, as they are unobservable. Hence, 
we only estimate the AOA and AOD at the RIS center using a FF approximation.  
Based on  \eqref{eq:RISreponses}, 
we can write FF approximation of $\mathbf{b}$ as
\begin{align}\label{eq:FF-approx}
[\mathbf{b}]_m =  e^{-\jmath \frac{2\pi}{\lambda}m \Delta \left(\sin{\left(\theta_\text{tx}\right)}+\sin{\left(\theta_\text{rx}\right)} \right)},
\end{align}
where $\theta_\text{tx}\triangleq \tilde{\theta}_\text{tx}+\alpha$ and $\theta_\text{rx}\triangleq \tilde{\theta}_\text{rx}+\alpha$ are AOA and AOD at the center of the RIS, respectively (see Fig.~\ref{System Setup}). Given  \eqref{eq:FF-approx}, we can estimate 
\begin{align}\label{eq:omega_def}
\omega \triangleq \sin{\left(\tilde{\theta}_\text{tx}+\alpha\right)}+\sin{\left(\tilde{\theta}_\text{rx}+\alpha\right)}.    
\end{align}
 Thus, considering \eqref{eq:sum-subcar}--\eqref{eq:omega_def}, we can rewrite \eqref{eq:sum-subcar} as 
\begin{align}\label{y_FF}
\mathbf{y}_r = N_c g_r \sqrt{P_t} \mathbf{\Gamma}^\top \mathbf{b\left(\omega\right)}+ \mathbf{w}_t.
\end{align}
Here, for each value of $\omega$, the value of $g_r$ can be estimated by  $\hat{g}_r (\omega) = \mathbf{b}^{\mathsf{H}}\left(\omega\right)\boldsymbol{\Gamma}^* \mathbf{y}_r / (N_c  \sqrt{P_t}\Vert \mathbf{b}^{\mathsf{H}}\left(\omega\right)\boldsymbol{\Gamma}^* \Vert^2)$. Hence, the estimation for $\omega$ can be obtained as 
\begin{align}\label{eq:MLE-omega}
\hat{\omega} = \arg\min_{\omega} \Vert \mathbf{y}_r - N_c  \sqrt{P_t} \hat{g}_r\left(\omega\right) \mathbf{\Gamma}^\top \mathbf{b} \left(\omega\right)\Vert^2,
\end{align}
which can be solved via a simple 1D line search. 
\subsubsection{RIS Position and Orientation Estimation}
 \begin{figure*}[t!]
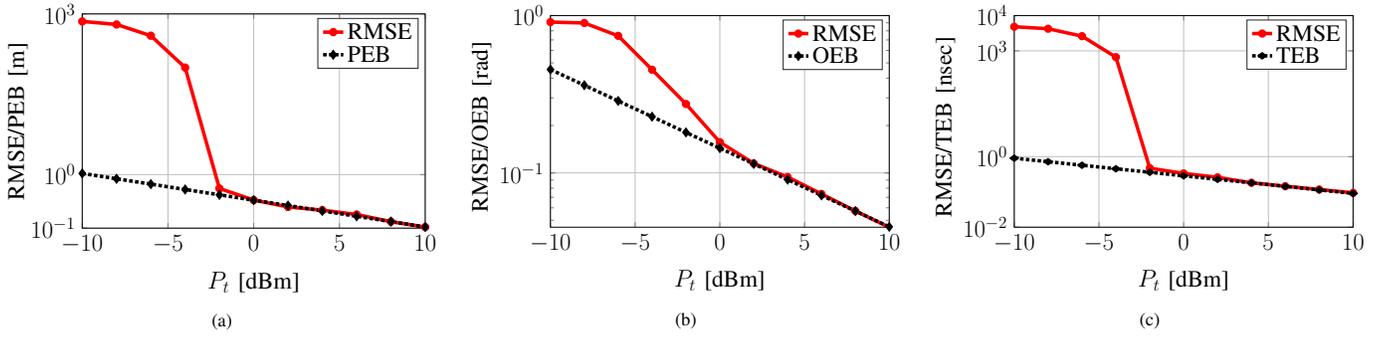

\begin{subfigure}[h!]{0.32\textwidth}
    \centering
    \includegraphics[width=\linewidth]{Figures/POS_SNR.tikz}
    \caption{}
    \label{fig: POS_SNR}
\end{subfigure}
\hfill
\begin{subfigure}[h!]{0.32\textwidth}
    \centering
    \includegraphics[width=\linewidth]{Figures/OEB_SNR.tikz}
    \caption{}
    \label{fig: OEB_SNR}
\end{subfigure}
\hfill
\begin{subfigure}[h!]{0.32\textwidth}
    \centering
    \includegraphics[width=\linewidth]{Figures/tau_SNR.tikz}
    \caption{}
    \label{fig: tau_SNR}
\end{subfigure}
\caption{The evaluation of the RMSE of the $\mathbf{p}_{\text{ris}}$, $\mathbf{\alpha}$, and $\mathbf{\tau}$ versus $P_t$. The simulation parameters are given in Table \ref{table: tab1}: (a) RMSE of the RIS position (meter) and PEB versus $P_t$; (b) RMSE of the RIS orientation (rad) and OEB versus $P_t$ ; (c) RMSE of the $\tau$ (nsec) and TEB versus $P_t$.}
\label{fig:PEB_OEB_PT}
\end{figure*}
Based on the  $\hat{\tau}$, the RIS is constrained to lie on the ellipse defined by $\Vert \mathbf{p}_{\text{ris}} - \mathbf{p}_{\text{tx}}\Vert + \Vert\mathbf{p}_{\text{ris}} - \mathbf{p}_{\text{rx}} \Vert = c\hat{\tau}$. Thus a possible solution of RIS position can be found by parameterizing this ellipse a function of $\nu\in [0,2\pi)$ as
\begin{align}\label{eq:ellipse}
\mathbf{p}_{\text{ris}}\left(\nu\right) = \mathbf{R}_{\beta} \mathbf{A}_{\nu} \mathbf{e}+ \mathbf{c}_e. 
\end{align}
Here, $\mathbf{c}_e = (\mathbf{p}_{\text{rx}}+\mathbf{p}_{\text{tx}})/2$ is the center of the ellipse, $\mathbf{R}_{\beta}$ is a rotation matrix with angle $\beta$, where $\beta$ is a angle between positive $x$-axis and the line between TX and RX.  $\mathbf{A}_{\nu} \triangleq \text{diag}(\left[\cos{\nu} , \sin{\nu}\right]^\top)$ and $\mathbf{e} = [ c\hat{\tau}/2 , \sqrt{a^2 - \Vert \mathbf{p}_{\text{tx}} - \mathbf{c}_e \Vert^2}]^\top$. 
Furthermore, we can formulate the $\tilde{\theta}_\text{tx}$ and $\tilde{\theta}_\text{rx}$ a function of $\nu$ as
\begin{align}\label{eq:theta_r}
  \tilde{\theta}_\text{rx} \left( \nu \right) &= \text{atan2}\left(y_{\text{n}_\text{r}},x_{\text{n}_\text{r}}\right)-\frac{\pi}{2},\\
\label{eq:theta_t}
  \tilde{\theta}_\text{tx} \left( \nu \right) &= \text{atan2}\left(y_{\text{n}_\text{t}},x_{\text{n}_\text{t}}\right)-\frac{\pi}{2},
\end{align}
where $ [x_{\text{n}_\text{r}},y_{\text{n}_\text{r}}]^\top = (\mathbf{p}_{\text{rx}}-\mathbf{p}_{\text{ris}}(\nu))$ and $ [x_{\text{n}_\text{t}},y_{\text{n}_\text{t}}]^\top = (\mathbf{p}_{\text{tx}}-\mathbf{p}_{\text{ris}}(\nu))$.

Based on the definition of $\omega$, $\hat{\omega}$, (\ref{eq:theta_r}) and (\ref{eq:theta_t}), $\alpha$ is also obtained as a function of $\nu$ 
\begin{align}\label{eq:alpha_nu}
   \alpha \left( \nu \right) = \arcsin\left(\frac{\hat{\omega}}{2 \cos{\theta_{\text{tr}}^-}} \right) - \theta_{\text{tr}}^+ \quad , 
\end{align}
where $\theta_{\text{tr}}^+ \triangleq(\tilde{\theta}_\text{tx} + \tilde{\theta}_\text{rx}) /2$  and $\theta_{\text{tr}}^- \triangleq(\tilde{\theta}_\text{tx} - \tilde{\theta}_\text{rx}) /2$. Hence, we can estimate  $\nu$  as
\begin{align}\label{eq:MLE_nu}
  \hat{\nu} &= \arg\min_{\nu} \Vert \mathbf{Y} - \hat{g}_r\left(\nu\right) \sqrt{P_t}\mathbf{d}\left(\hat{\tau}\right) \mathbf{b\left(\mathbf{p}_{\text{ris}}\left(\nu\right),\alpha \left( \nu \right)\right)}^\top \boldsymbol{\Gamma}\Vert^2, \end{align}
where $\hat{g}_r(\nu)  = \mathbf{h}^{\mathsf{H}}\left(\nu\right) \mathbf{y} / (\sqrt{P_t}\Vert \mathbf{h}\left(\nu\right)\Vert^2)$,  $\mathbf{y} = \mathbf{Y}^\top \mathbf{1}_{N_c}$ and  $\mathbf{h}(\nu) \triangleq (\mathbf{d}(\hat{\tau}) \mathbf{b(\mathbf{p}_{\text{ris}}(\nu),\alpha ( \nu ))}^\top \boldsymbol{\Gamma})^\top \mathbf{1}_{N_c}$,  $\mathbf{b(\mathbf{p}_{\text{ris}}(\nu),\alpha ( \nu ))}$ is defined based on \eqref{eq:at} and \eqref{eq:ar}. Note that \eqref{eq:MLE_nu} can be solved through a line search. Then, the position and orientation of the RIS can be obtained via \eqref{eq:ellipse}--\eqref{eq:MLE_nu}. This solution is applied as the initial guess for quasi-Newton method to solve \eqref{eq:MLE}.  

In summary, the RIS location can be recovered by several 1D line searches, which leads to very low complexity. 



\section{Simulation Results}

In this section, we evaluate the proposed estimator and compare it to the corresponding bound. To do so, we compare the root mean square error (RMSE) of the estimated parameters  (e.g, RIS position) with the derived CRB. 

\subsection{Simulation Parameters}
 
To evaluate the performance of the proposed estimator, we average over $500$ noise realizations. We assume that the global coordinate system is set to be aligned with RX coordinate system. The phase profile at the RIS elements is randomly drawn from the uniform distribution $[0,2\pi)$. The number of RIS elements is set to $M=64$. The rest of the simulation parameters are given in Table \ref{table: tab1}. 
 
 \begin{table}[b!]
\caption{Simulation Parameters}
\begin{center}
\begin{tabular}{l c c} 
 \hline \hline
 Parameter & Symbol & Value  \\  
 \hline\hline
 Wavelength & $\lambda$ & $1 ~\text{cm}$\\
 RIS element spacing & $\Delta$ & $0.5 ~\text{cm}$ \\Light speed & $c$ & $3\times 10^{8}~ \text{m}/\text{sec}$\\
 Number of sub-carriers & $N_c$& $500$\\
 TX antenna gain & $G_t$& $2$\\
 RX antenna gain & $G_r$& $2$\\
 Number of transmissions & $T$& $50$\\
 Sub-carrier spacing & $\Delta f$& $120~\text{kHz}$\\
 Noise PSD&$N_0$& $-174~\text{dBm/Hz}$\\ RX's noise figure& $n_f$ & $8~\text{dB}$\\Noise variance&$\sigma^2 = n_f N_0 N_c \Delta f$&$-88~\text{dBm}$\\ IFFT Size & $N_F$& $4096$\\ RX position& $\mathbf{p}_{\text{rx}}$& $\left[2,2\right]^\top$ \\TX position& $\mathbf{p}_{\text{tx}}$& $\left[0,2\right]^\top$ \\RIS position& $\mathbf{p}_{\text{ris}}$& $\left[0,0\right]^\top$\\
 RIS orientation& $\alpha$& $\pi/6~\text{rad}$\\
 \hline\hline
\end{tabular}
\label{table: tab1}
\end{center}
\end{table}
\subsection{Results and Discussion}
 In Fig.~\ref{fig:PEB_OEB_PT} , we study the effect of the transmitted power on the performance of the proposed estimator. As expected, the RMSE of the $\mathbf{p}_{\text{ris}}$, $\alpha$, and $\tau$ are decreasing functions of the transmitted power $P_t$ and 
 attains the CRB when $P_t\ge 0$ dBm. 

\begin{figure*}[h!]
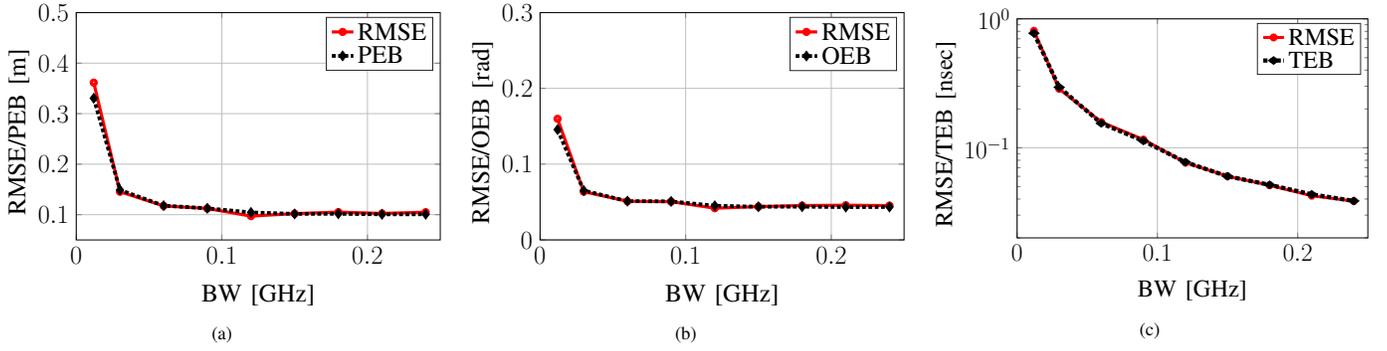

\begin{subfigure}[h!]{0.32\textwidth}
    \centering
    \includegraphics[width=\linewidth]{Figures/PEB_BW.tikz}
    \caption{}
    \label{fig: PEB_BW}
\end{subfigure}
\hfill
\begin{subfigure}[h!]{0.32\textwidth}
    \centering
    \includegraphics[width=\linewidth]{Figures/OEB_BW.tikz}
    \caption{}
    \label{fig: OEB_BW}
\end{subfigure}
\hfill
\begin{subfigure}[h!]{0.32\textwidth}
    \centering
    \includegraphics[width=\linewidth]{Figures/tau_BW.tikz}
    \caption{}
    \label{fig: tau_BW}
\end{subfigure}
\caption{The comparison of the RMSE of the $\mathbf{p}_{\text{ris}}$ and $\mathbf{\alpha}$ with CRB with respect to the system BW. $P_t = 10~\text{dBm}$, and the rest of  simulation parameters are given in Table \ref{table: tab1}: (a) RMSE of the RIS position (meter) and PEB versus BW; (b) RMSE of the RIS orientation (rad) and OEB versus BW; (c) RMSE of $\tau$ (nsec) and TEB versus BW.}
\label{fig:EB_BW}
\end{figure*}

Next, we investigate the effect of the system BW on the RMSE of the RIS localization, shown in Fig.~\ref{fig:EB_BW}. To this end, we increase the number of sub-carriers while the frequency spacing and transmit power are fixed. From Fig.~\ref{fig: tau_BW}, it is clear that as the BW increases, the RMSE of $\tau$ decreases. 
Regrading the proposed method, $\tau$ estimation plays a key role on the other stages of the proposed estimator. The more accurate $\tau$ estimate, the more accurate $\omega$ and RIS positioning. This can be seen in Fig.~\ref{fig: PEB_BW} and Fig.~\ref{fig: OEB_BW}:  with better $\tau$ estimation, the  ellipse equation \eqref{eq:ellipse} becomes more accurate. 
Since the RIS location is on this ellipse, the RIS positioning accuracy will be enhanced. On the other hand, the accuracy of the $\omega$ estimate directly depends on the $\tau$ estimate (see \eqref{eq:remove_d}--\eqref{eq:MLE-omega}). The $\omega$ estimate provides information about the RIS orientation. Therefore, with finer $\tau$ and $\omega$ estimation, the estimator achieves higher accuracy in RIS location and orientation estimate. This is the effect of the BW increase. For bandwidths larger than 100 MHz, we observe that the performance is limited by the the estimation of $\omega$. 
 
\begin{figure}[h!]
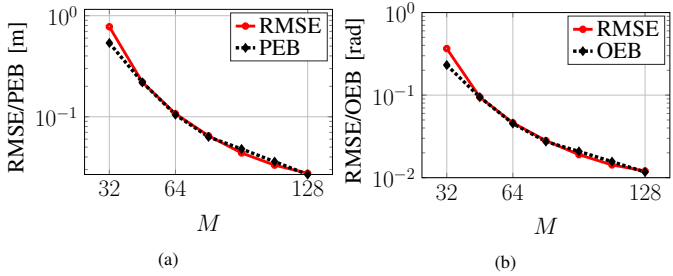

\begin{subfigure}[h!]{0.241\textwidth}
    \centering
    \includegraphics[width=\linewidth]{Figures/PEB_RIS_elem.tikz}
    \caption{}
    \label{fig: PEB_size}
\end{subfigure}
\hfill
\begin{subfigure}[h!]{0.241\textwidth}
    \centering
    \includegraphics[width=\linewidth]{Figures/OEB_RIS_elem.tikz}
    \caption{}
    \label{fig: OEB_size}
\end{subfigure}
\caption{The comparison of the RMSE of the $\mathbf{p}_{\text{ris}}$ and $\mathbf{\alpha}$ with CRB with respect to the RIS size. $P_t = 10\text{dBm}$, and the rest of the simulation parameters are given in Table \ref{table: tab1}: (a) RMSE of the RIS position (meter) and PEB versus the RIS size; (b) RMSE of the RIS orientation (rad) and OEB versus the RIS size.}
\label{fig:PEB_OEB_size}
\end{figure}

 Now, we analyze the impact of the RIS size on the RIS localization as shown in Fig.~\ref{fig:PEB_OEB_size}. As it is shown in  Fig.~\ref{fig: PEB_size} and Fig.~\ref{fig: OEB_size}, the RMSE of the RIS localization reduces as the number of the RIS elements grows. One can also see that the proposed method can localize RIS with sub-centimeter accuracy when the number of RIS elements reaches $128$. This observation is due to two compounding effects: a larger RIS size leads to higher SNR, and the larger RIS size leads to a more pronounced NF effect. 

\begin{figure}[h!]
\begin{subfigure}[t]{0.24\textwidth}
    \centering
    \includegraphics[width=\linewidth]{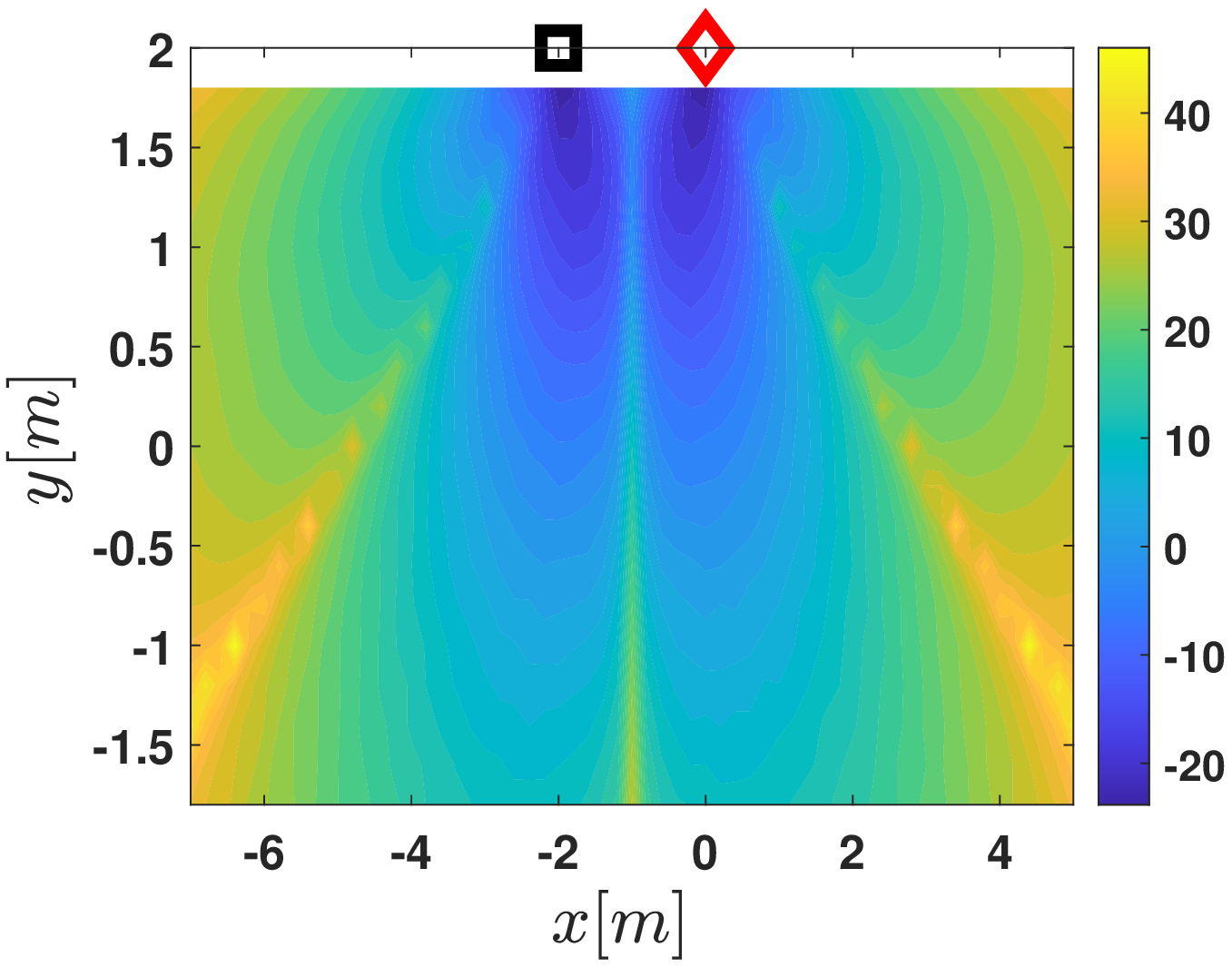}
    \caption{}
    \label{fig:PEB_LOC_0deg}
\end{subfigure}
\hfill
\begin{subfigure}[t]{0.24\textwidth}
    \centering
    \includegraphics[width=\linewidth]{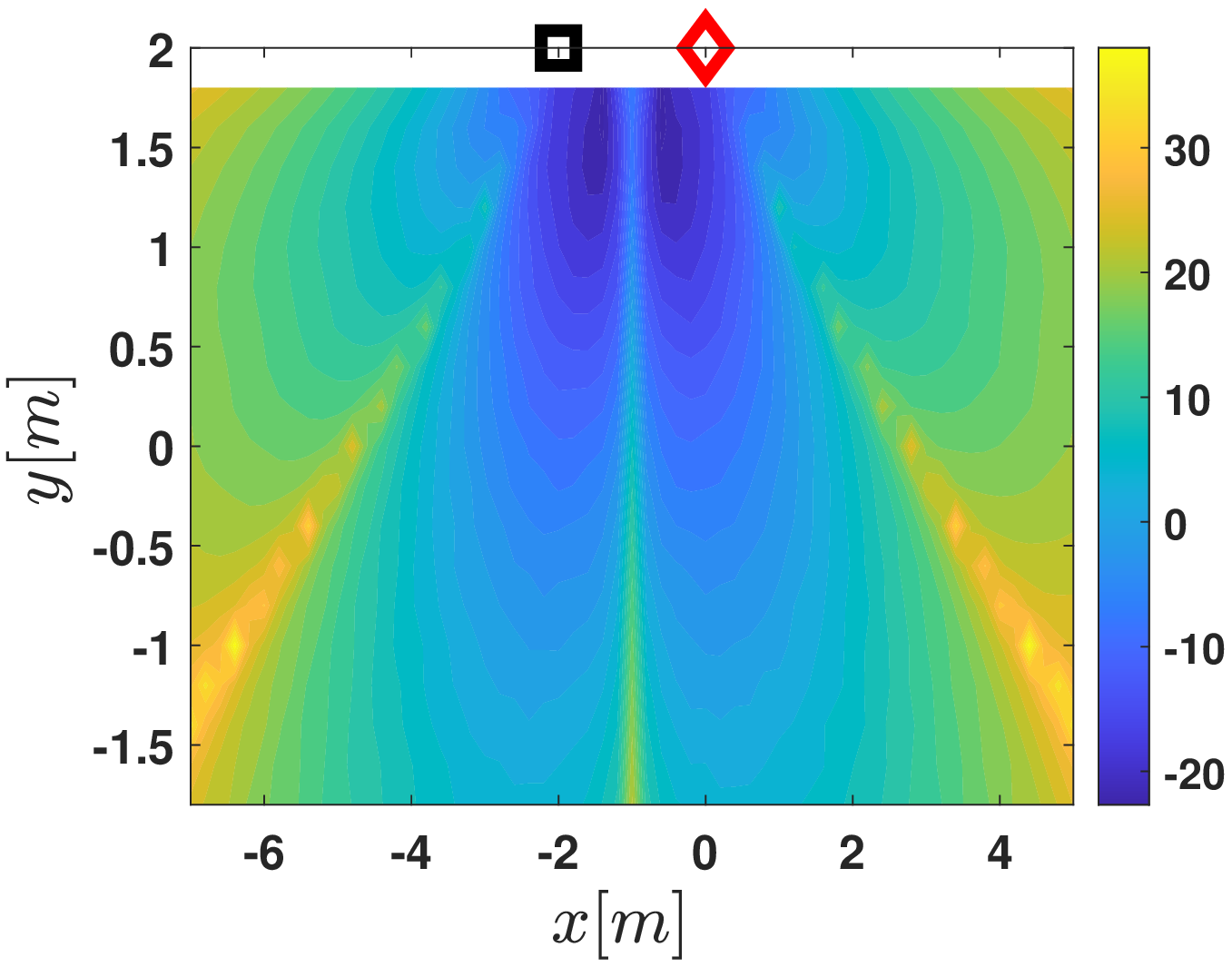}
    \caption{}
    \label{fig:OEB_LOC_0deg}
\end{subfigure}
\hfill
\begin{subfigure}[t]{0.24\textwidth}
    \centering
    \includegraphics[width=\linewidth]{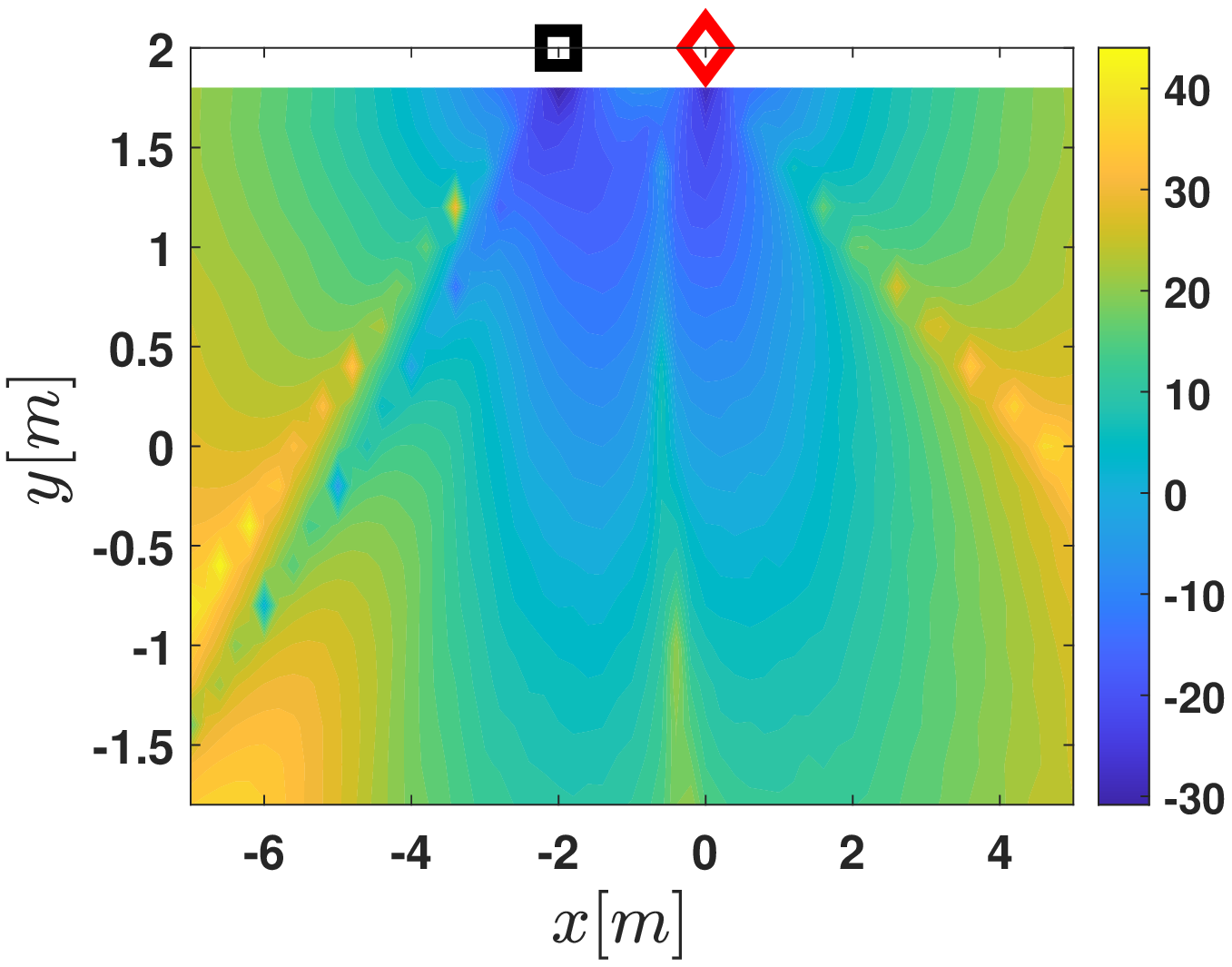}
    \caption{}
    \label{fig:PEB_LOC}
\end{subfigure}
\begin{subfigure}[t]{0.24\textwidth}
    \centering
    \includegraphics[width=\linewidth]{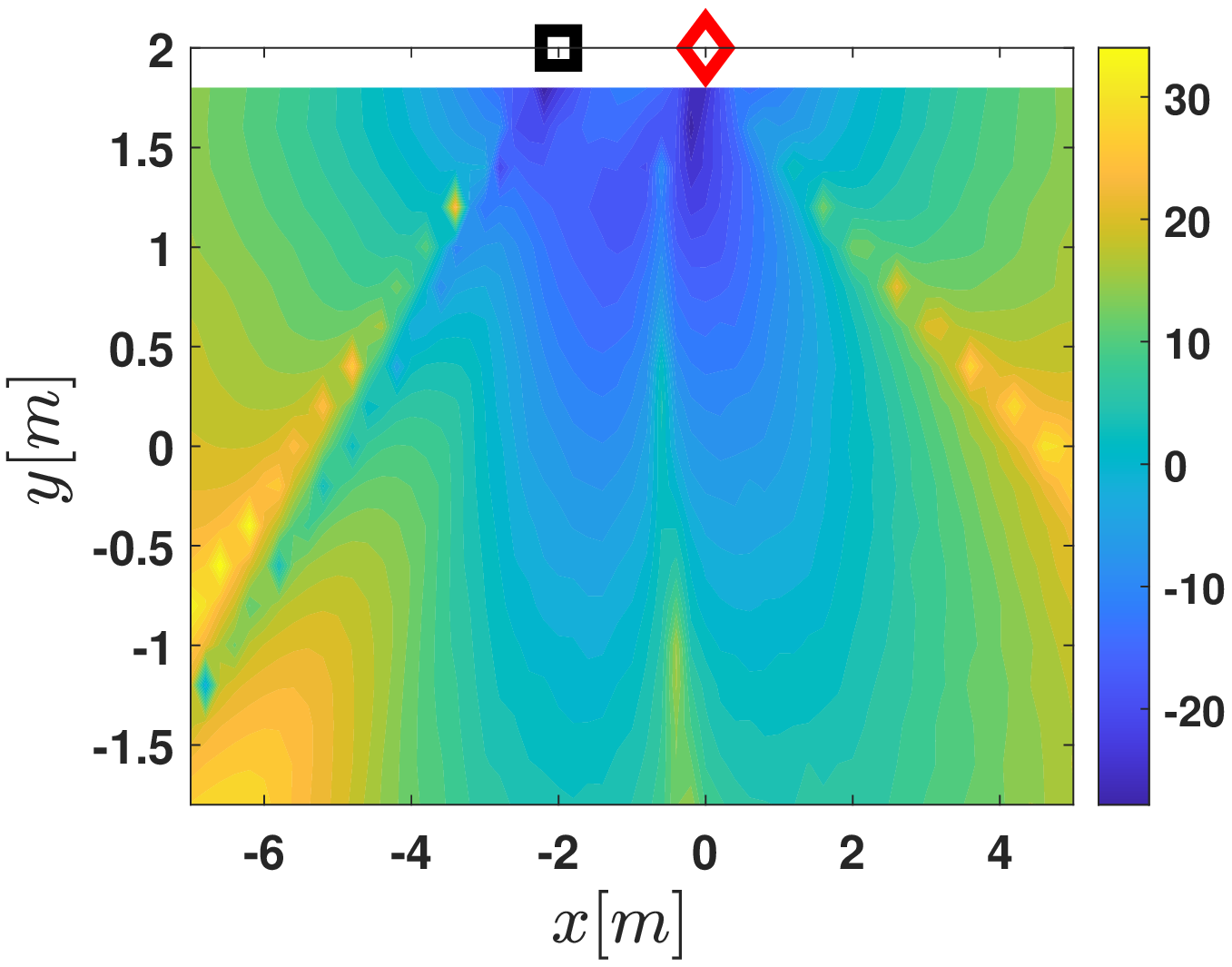}
  \caption{}
   \label{fig:OEB_LOC}
\end{subfigure}
\caption{The contour plots of PEB and OEB versus the RIS location for one random RIS phase profile and $P_t = 10\,\text{dBm}$. TX and RX positions are marked by red diamond and black square, respectively: (a) $10\log10(\text{PEB})$ (meter), $\alpha = 0$ (rad);(b) $10\log10(\text{OEB})$ (rad), $\alpha = 0$ (rad);(c) $10\log10(\text{PEB})$ (meter), $\alpha = \pi/6$ (rad); (d) $10\log10(\text{OEB})$ (rad), $\alpha = \pi/6$.}
\label{fig:contour}
\end{figure}

Finally, we assess the RIS localization coverage and performance through contour plots of PEB and OEB. Fig.~\ref{fig:contour} illustrates the contour plots of PEB and OEB when the $x$ and $y$ coordinates of the RIS are varied while its orientation is fixed. To evaluate the effect of the RIS orientation, we depict contour plots for two cases: (i) $\alpha = 0 $  (rad); (ii) $\alpha = \pi/6 $ (rad). 
One can see, in general, that high accurate location and orientation can be obtained when the RIS is close to the RX or TX, due to high SNR. However, several blind areas exist (e.g., $[-1, -0.5]^\top$ and $[-6, -1]^\top$ in Fig.~\ref{fig:PEB_LOC_0deg}).
For $\alpha = 0$ (rad), both location and orientation estimate degrade as the RIS approaches the bisector of the line segment between TX and RX. 
Furthermore, we can also see two other areas where RIS localization has a high error. When the RIS rotates (Fig.~\ref{fig:PEB_LOC} and Fig.~\ref{fig:OEB_LOC}) the mentioned lines rotates accordingly.

\section{Conclusion}
In this work, we have proposed a low-complexity bi-static sensing estimator to determine the  location and orientation of RIS in NF with the help of a single-antenna transmitter and receiver. We have also derived the CRBs on the RIS position and orientation estimate error.  To show the efficiency of the estimator, the RMSE of the estimations has been compared with the CRBs. Simulation results have verified the estimator's accuracy such that its RMSE attains the CRB. To gain insight into the RIS localization problem, we investigated the influential factors like transmitted power, the number of RIS elements, BW, and RIS position and orientation concerning the TX/RX. Results show that increasing the value of these factors enhances the RIS localization accuracy. In future work, we would like to study 3D RIS localization.
\vspace{-2mm}
\section*{Acknowledgment}
This work has been funded in part by Academy of Finland Profi-5 (n:o 326346) and ULTRA (n:o 328215) projects, and the EU H2020 RISE-6G project under
grant 101017011.

\section*{Appendix}
\subsection{Derivation of $\mathbf{J}$}
In this appendix, we drive the elements of the matrix $\mathbf{J}$.  To this end, we derive $\partial [\boldsymbol{\mu}]_{t,n_c}/\partial\boldsymbol{\eta}$ according to (\ref{eqn:FIM WB}) as $\partial [\boldsymbol{\mu}]_{t,n_c}/\partial\rho=   e^{\jmath\phi} \sqrt{P_t} e^{-\jmath 2 \pi\tau n_c \Delta f} \mathbf{b}^\top  \boldsymbol{\gamma}(t)$,  $\partial [\boldsymbol{\mu}]_{t,n_c}/\partial\phi =  \jmath \rho  e^{\jmath\phi} \sqrt{P_t} e^{-\jmath 2 \pi\tau n_c \Delta f} \mathbf{b}^\top  \boldsymbol{\gamma}(t)$, $\partial [\boldsymbol{\mu}]_{t,n_c}/\partial\boldsymbol{\tau}= -\jmath 2\pi n_c \Delta f \rho e^{\jmath\phi}  \sqrt{P_t} e^{-\jmath 2 \pi\tau n_c \Delta f}  \mathbf{b}^\top \boldsymbol{\gamma}(t)$, and 
\begin{align}\label{eq:d-pris}
&\frac{ \partial [\boldsymbol{\mu}]_{t,n_c}}{\partial\boldsymbol{p}_{\text{ris}}} = \rho e^{\jmath\phi}\sqrt{P_t} e^{-\jmath 2 \pi\tau n_c \Delta f} \mathbf{E}^\top  \boldsymbol{\gamma}(t) \end{align}
\vspace{.02mm}
\begin{align}\label{eq:d-alpha}
&\frac{ \partial [\boldsymbol{\mu}]_{t,n_c}}{\partial\alpha} = \rho  e^{\jmath\phi} \sqrt{P_t} e^{-\jmath 2 \pi\tau n_c \Delta f} \mathbf{v}^\top \boldsymbol{\gamma}(t) \end{align}
where $\mathbf{E} \triangleq \partial\mathbf{b}/\partial \mathbf{p}_{\text{ris}} $,$\mathbf{v} \triangleq \partial \mathbf{b}/\partial \alpha$, and we have 
\begin{align}\label{eq:h elem}
&\left[\mathbf{E}\right]_{:,m}= -\jmath \frac{2\pi}{\lambda} \left( \boldsymbol{\kappa}_m^t + \boldsymbol{\kappa}_m^r - \boldsymbol{\kappa}_o^t - \boldsymbol{\kappa}_o^r      \right) \left[\mathbf{b} \right]_m \end{align}
where $\boldsymbol{\kappa}_m^t \triangleq (((m-(M-1)/2 )\Delta-\mathbf{p}_{\text{tx}})/R_m^t$, $\boldsymbol{\kappa}_m^r \triangleq (((m-(M-1)/2 )\Delta-\mathbf{p}_{\text{rx}})/R_m^r$, $\boldsymbol{\kappa}_o^t \triangleq (\mathbf{p}_{\text{ris}}-\mathbf{p}_{\text{rx}})/R_o^t$, and $\boldsymbol{\kappa}_o^r \triangleq (\mathbf{p}_{\text{ris}}-\mathbf{p}_{\text{rx}})/R_o^r$,
\begin{align}\label{eq:v elem}
&\left[\mathbf{v}\right]_m  = -\jmath \frac{2\pi}{\lambda} \left( \boldsymbol{\psi}_m^t + \boldsymbol{\psi}_m^r       \right) \left[\mathbf{b} \right]_m ,  \end{align}
where $\boldsymbol{\psi}_m^t \triangleq ((\mathbf{p}_m-\mathbf{p}_{\text{tx}})/R_m^t) \mathbf{ R}^{\prime}_{\alpha}(((m-(M-1)/2 )\Delta) $, $\boldsymbol{\psi}_m^r \triangleq ((\mathbf{p}_m-\mathbf{p}_{\text{rx}})/R_m^r) \mathbf{ R}^{\prime}_{\alpha}(((m-(M-1)/2 )\Delta)$, and  $\mathbf{ R}^{\prime}_{\alpha}\triangleq \partial \mathbf{R}_{\alpha}/\partial \alpha$.


\subsection{Derivation of the  Jacobian Matrix $\mathbf{T}$}
Based on  \eqref{eqn:Jacobian} and the relation between $\tau$, $\phi$, $\rho$ and $\mathbf{p}_{\text{ris}}$, the Jacobian matrix elements can be written as $\left[\mathbf{T}\right]_{1:2,1:2} = \partial \left[\rho , \phi\right]^\top/\partial \left[\rho , \phi\right]^\top = \mathbf{I}_{2\times2}$, and 
\begin{align}
\left[\mathbf{T}\right]_{3,3:4} = \frac{\partial \tau}{\partial \mathbf{p}_{\text{ris}}}= \frac{1}{c}  \left( \mathbf{d}_{tx}^\top + \mathbf{d}_{rx}^\top\right) 
\end{align}
where $\mathbf{d}_{tx} = (\mathbf{p}_{\text{ris}} - \mathbf{p}_{\text{tx}}) / \Vert\mathbf{p}_{\text{ris}} - \mathbf{p}_{\text{tx}}\Vert$ and $\mathbf{d}_{rx} = (\mathbf{p}_{\text{ris}} - \mathbf{p}_{\text{rx}})/\Vert\mathbf{p}_{\text{ris}} - \mathbf{p}_{\text{rx}}\Vert$. 
Besides, we have $\left[\mathbf{T}\right]_{4:5,3:4} = \partial \mathbf{p}_{\text{ris}}/\partial \mathbf{p}_{\text{ris}}=  \mathbf{I}_{2\times2}$ and $\left[\mathbf{T}\right]_{6,5} =\partial\alpha /\partial\alpha= 1$. 
The other elements of the matrix $\mathbf{T}$ are zero.

\balance 
\bibliographystyle{IEEEtran}
\bibliography{ref.bib}

\begin{thebibliography}{10}
\providecommand{\url}[1]{#1}
\csname url@samestyle\endcsname
\providecommand{\newblock}{\relax}
\providecommand{\bibinfo}[2]{#2}
\providecommand{\BIBentrySTDinterwordspacing}{\spaceskip=0pt\relax}
\providecommand{\BIBentryALTinterwordstretchfactor}{4}
\providecommand{\BIBentryALTinterwordspacing}{\spaceskip=\fontdimen2\font plus
\BIBentryALTinterwordstretchfactor\fontdimen3\font minus
  \fontdimen4\font\relax}
\providecommand{\BIBforeignlanguage}[2]{{%
\expandafter\ifx\csname l@#1\endcsname\relax
\typeout{** WARNING: IEEEtran.bst: No hyphenation pattern has been}%
\typeout{** loaded for the language `#1'. Using the pattern for}%
\typeout{** the default language instead.}%
\else
\language=\csname l@#1\endcsname
\fi
#2}}
\providecommand{\BIBdecl}{\relax}
\BIBdecl

\bibitem{wymeersch2020radio}
H.~Wymeersch, J.~He, B.~Denis, A.~Clemente, and M.~Juntti, ``Radio localization
  and mapping with reconfigurable intelligent surfaces: Challenges,
  opportunities, and research directions,'' \emph{IEEE Vehicular Technology
  Magazine}, vol.~15, no.~4, pp. 52--61, 2020.

\bibitem{basar2019wireless}
E.~Basar, M.~Di~Renzo, J.~De~Rosny, M.~Debbah, M.-S. Alouini, and R.~Zhang,
  ``Wireless communications through reconfigurable intelligent surfaces,''
  \emph{IEEE access}, vol.~7, pp. 116\,753--116\,773, 2019.

\bibitem{yuan2021frequency}
J.~Yuan, E.~De~Carvalho, R.~J. Williams, E.~Bj{\"o}rnson, and P.~Popovski,
  ``Frequency-mixing intelligent reflecting surfaces for nonlinear wireless
  propagation,'' \emph{IEEE Wireless Communications Letters}, 2021.

\bibitem{strinati2021reconfigurable}
E.~C. Strinati, G.~C. Alexandropoulos, H.~Wymeersch, B.~Denis,
  V.~Sciancalepore, R.~D'Errico, A.~Clemente, D.-T. Phan-Huy, E.~De~Carvalho,
  and P.~Popovski, ``Reconfigurable, intelligent, and sustainable wireless
  environments for 6{G} smart connectivity,'' \emph{IEEE Communications
  Magazine}, vol.~59, no.~10, pp. 99--105, 2021.

\bibitem{basar2020reconfigurable}
E.~Basar, ``Reconfigurable intelligent surface-based index modulation: A new
  beyond mimo paradigm for 6{G},'' \emph{IEEE Transactions on Communications},
  vol.~68, no.~5, pp. 3187--3196, 2020.

\bibitem{chen2021tutorial}
H.~Chen, H.~Sarieddeen, T.~Ballal, H.~Wymeersch, M.-S. Alouini, and T.~Y.
  Al-Naffouri, ``A tutorial on terahertz-band localization for {6G}
  communication systems,'' \emph{Accepted for publication in {IEEE} Commun.
  Surveys Tuts. arXiv preprint arXiv:2110.08581}, 2022.

\bibitem{xing2021location}
Z.~Xing, R.~Wang, X.~Yuan, and J.~Wu, ``Location-aware beamforming design for
  reconfigurable intelligent surface aided communication system,'' in
  \emph{2021 IEEE/CIC International Conference on Communications in China
  (ICCC)}, 2021.

\bibitem{keykhosravi2021siso}
K.~Keykhosravi, M.~F. Keskin, G.~Seco-Granados, and H.~Wymeersch, ``{SISO
  RIS}-enabled joint 3{D} downlink localization and synchronization,'' in
  \emph{IEEE International Conference on Communications}, 2021.

\bibitem{abu2021near}
Z.~Abu-Shaban, K.~Keykhosravi, M.~F. Keskin, G.~C. Alexandropoulos,
  G.~Seco-Granados, and H.~Wymeersch, ``Near-field localization with a
  reconfigurable intelligent surface acting as lens,'' in \emph{ICC 2021-IEEE
  International Conference on Communications}, 2021.

\bibitem{huang2022near}
S.~Huang, B.~Wang, Y.~Zhao, and M.~Luan, ``Near-field {RSS}-based localization
  algorithms using reconfigurable intelligent surface,'' \emph{IEEE Sensors
  Journal}, 2022.

\bibitem{he2020adaptive}
J.~He, H.~Wymeersch, T.~Sanguanpuak, O.~Silv{\'e}n, and M.~Juntti, ``Adaptive
  beamforming design for mmwave {RIS}-aided joint localization and
  communication,'' in \emph{2020 IEEE Wireless Communications and Networking
  Conference Workshops (WCNCW)}, 2020.

\bibitem{hu2020location}
X.~Hu, C.~Zhong, Y.~Zhang, X.~Chen, and Z.~Zhang, ``Location information aided
  multiple intelligent reflecting surface systems,'' \emph{IEEE Transactions on
  Communications}, vol.~68, no.~12, pp. 7948--7962, 2020.

\bibitem{keykhosravi2021semi}
K.~Keykhosravi, M.~F. Keskin, S.~Dwivedi, G.~Seco-Granados, and H.~Wymeersch,
  ``Semi-passive 3{D} positioning of multiple {RIS}-enabled users,'' \emph{IEEE
  Transactions on Vehicular Technology}, vol.~70, no.~10, pp. 11\,073--11\,077,
  2021.

\bibitem{sherman1962properties}
J.~Sherman, ``Properties of focused apertures in the fresnel region,''
  \emph{IRE Transactions on Antennas and Propagation}, vol.~10, no.~4, pp.
  399--408, 1962.

\bibitem{tang2020wireless}
W.~Tang, M.~Z. Chen, X.~Chen, J.~Y. Dai, Y.~Han, M.~Di~Renzo, Y.~Zeng, S.~Jin,
  Q.~Cheng, and T.~J. Cui, ``Wireless communications with reconfigurable
  intelligent surface: Path loss modeling and experimental measurement,''
  \emph{IEEE Transactions on Wireless Communications}, vol.~20, no.~1, pp.
  421--439, 2020.

\bibitem{friedlander2019localization}
B.~Friedlander, ``Localization of signals in the near-field of an antenna
  array,'' \emph{IEEE Transactions on Signal Processing}, vol.~67, no.~15, pp.
  3885--3893, 2019.

\bibitem{kay1993fundamentals}
S.~M. Kay, \emph{Fundamentals of statistical signal processing: estimation
  theory}.\hskip 1em plus 0.5em minus 0.4em\relax Prentice-Hall, Inc., 1993.

\end{thebibliography}


\end{document}